\documentclass[iop]{emulateapj}
\usepackage{amsmath}
\usepackage{amsfonts}
\usepackage{amssymb}
\usepackage{graphicx}
\usepackage{enumitem}
\begin{document}

\title{WASP-104b is Darker than Charcoal}
\author{T.~Mo\v{c}nik, C.~Hellier, and J.~Southworth}
\affil{Astrophysics Group, Keele University, Staffordshire, ST5 5BG, UK}
\email{t.mocnik@keele.ac.uk}
\shorttitle{WASP-104b is Darker than Charcoal}
\shortauthors{Mo\v{c}nik et al.}

\begin{abstract}
By analysing the \textit{K2} short-cadence data from Campaign 14 we detect phase-curve modulation in the light curve of the hot-Jupiter host star WASP-104. The ellipsoidal modulation is detected with high significance and in agreement with theoretical expectations, while Doppler beaming and reflection modulations are detected tentatively. We show that the visual geometric albedo is lower than 0.03 at 95\% confidence, making it one of the least-reflective planets found to date. The light curve also exhibits a rotational modulation, implying a stellar rotational period likely to be near 23 or 46 days. In addition, we refine the system parameters and place tight upper limits for transit timing and duration variations, starspot occultation events, and additional transiting planets.
\end{abstract}


\section{INTRODUCTION}

Planetary phase curves consist of four components: (1) reflection of starlight from the surface of the orbiting planet \citep{Jenkins03}; (2) the planet's thermal emission \citep{Charbonneau05}; (3) Doppler beaming caused by the orbital motion of the host star \citep{Loeb03}; and (4) ellipsoidal modulation caused by the rotation of the host star which is gravitationally distorted into an ellipsoid by the planet \citep{Pfahl08}. Additionally, transiting planets also produce secondary eclipses whenever their reflected and emitted light is blocked by the occulting star (e.g. \citealt{Angerhausen15}). Typical amplitudes of the individual phase-curve modulation components in planetary systems reported so far are of the order of a few tens of parts per million (ppm) at optical wavelengths (e.g. \citealt{Esteves13}). Detection of phase-curve modulations can reveal any non-transiting planets (e.g. \citealt{Millholland17}), can provide an independent determination of planet-to-star mass ratio, and enable a basic insight into the planetary atmospheric or surface characteristics such as the planetary albedo, day-night temperature contrast and the location offset of the hottest region from the sub-stellar point (e.g. \citet{Shporer17} and citations therein).

Theoretical atmospheric models suggest that cloud-free hot Jupiters have low geometric albedos at visual wavelengths due to strong and broad absorption lines of atomic Na and K (e.g. \citealt{Rowe08}). \citet{Heng13} have shown that there is no clear trend between the geometric albedo and the incident stellar flux and suggested that the correlation is hindered by the opacity effects in the planetary atmospheres, such as condensates or clouds, and atmospheric circulation. Expanding the sample of planets with known albedos to a wider variety of planetary systems and at different wavelengths is important for better understanding the underlying reflection mechanisms.

The \textit{K2} spacecraft \citep{Howell14} provides the community with high-precision long- (30\thinspace min) and short-cadence (1\thinspace min) photometry with nearly-continuous $\sim$80-day observing campaigns. This makes \textit{K2} well suited for the search of phase-curve modulations in visual wavelengths.

We present the analysis of the \textit{K2} short-cadence observations of WASP-104 \citep{Smith14}. Beside the detection of individual phase-curve modulation components, we also detect the rotational modulation, refine planetary system parameters, search for starspot occultation events, additional transiting planets, and transit-timing (TTVs) and transit-duration variations (TDVs).

WASP-104b is a transiting hot Jupiter in a 1.76-day circular orbit around a $V=11.1$ G8 main-sequence star \citep{Smith14}. The planet has a mass of 1.3\thinspace $M_{\rm Jup}$ and a radius of 1.1\thinspace $R_{\rm Jup}$. Unlike many other hot Jupiters, WASP-104b is not inflated. \citet{Smith14} also reported a non-detection of rotational modulation with an upper limit of 4\thinspace mmag at 95\% confidence.

\section{\textit{K2} OBSERVATIONS AND DATA REDUCTION}

WASP-104 was observed by \textit{K2} during the observing Campaign 14, which covered a time-span of 80 days between 2017 June 1 and 2017 August 19. We downloaded the short-cadence target pixel file from the Mikulski Archive for Space Telescopes (MAST) and performed a data reduction procedure as described in \citet{Mocnik16a} with PyRAF tools for Kepler (PyKE; \citealt{Still12}), optimized for short-cadence data.

We first defined a fixed and circular photometric extraction mask of 37 pixels, centered near the mean position of the target. The optimal mask size was chosen by trial and error as the best compromise between capturing as much starlight as possible and fewest possible background pixels. Choosing the mask too small resulted in larger residual systematics in the final reduced light curve and choosing the mask too large yielded higher white noise. Once the mask was defined, we extracted the light curve by summing the recorded flux values for each pixel within the extraction mask for every image in the target pixel file. The background was already subtracted as part of the Science Operations Center's calibration pipeline \citep{Quintana10}.

The main systematic errors present in the \textit{K2} light curves are the sawtooth-like artefacts caused by the pointing drift of the spacecraft. To correct for these artefacts we first removed any low-frequency variability by dividing the observed flux with the mean of the overlapping second-order polynomials with a 3-day window size, 0.3-day step size and a 3-$\sigma$ rejection threshold. After the flattening, we performed a self-flat-fielding (SFF) procedure presented in \citet{Mocnik16a}. In short, we used the Gaussian convolution to find a correlation between the measured flattened flux and the arclength of the spacecraft's drift. The SFF correction was split in 5-day time windows, outliers masked as 4-$\sigma$ outliers, and the width of the Gaussian kernel was chosen as 50 data points. All parameters were chosen based on trial and error to minimize the artefact residuals. We also masked planetary transits with a phase width of 0.045 to improve the rejection of data points for obtaining the SFF correlation near the beginning of ingress and end of egress. To remove the drift artefacts, we divided the flattened and normalized flux values with the measured correlation for each data point. This procedure removed virtually every trace of drift artefacts (see Figure~1) and improved the median 1-min photometric precision from 362\thinspace ppm before the SFF correction to 326\thinspace ppm after the correction. For comparison, the theoretical uncertainty propagation for similarly bright stars through the data processing pipeline of the \textit{Kepler} mission was $\sim$290\thinspace ppm \citep{Koch10}. The applied SFF procedure was highly effective firstly because the direction of spacecraft drifts was more consistent than on average in other observing campaigns, and secondly, because WASP-104 was placed in the central CCD module where the contribution of the spacecraft rotation to drift artefacts is smallest. Finally, we reintroduced the low-frequency modulations by multiplying the SFF-corrected flattened and normalized light curve with the same function as we used to flatten the light curve prior to SFF procedure. After rejecting the quality-flagged data points, such as thruster firing or cosmic ray events, we retained 113\thinspace 127 of the original 117\thinspace 030 data points present in the target pixel file. Figure~1 shows the binned, fluxed light curve before and after the SFF correction.

\begin{figure}
\includegraphics[width=8.5cm]{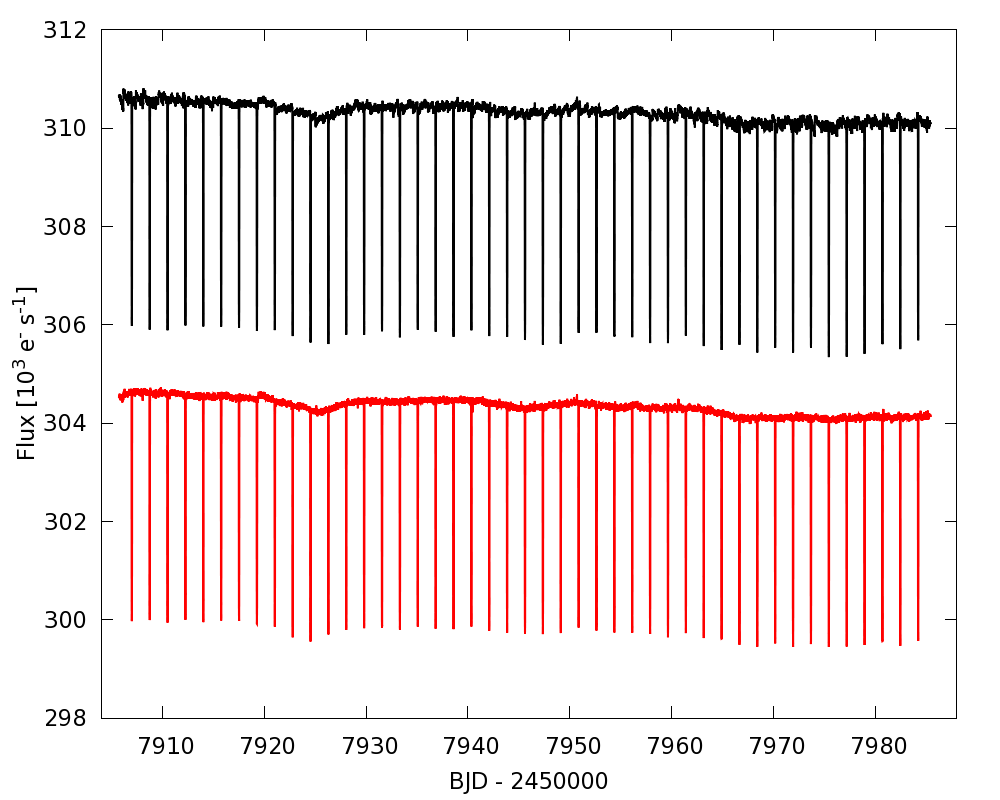}
\caption{Light curve of WASP-104 before (shown in black) and after the drift correction (red). Both light curves are shown with 10-min binning and contain 45 transits. The drift-corrected light curve is offset by \mbox{$-6\thinspace 000\thinspace\rm{e}^{-}\rm{s}^{-1}$} for clarity.}
\end{figure}

The corrected light curve in Figure~1 reveals not only the hot Jupiter's transits but also indicates the presence of a stellar rotational modulation (see Section~5) and a dropping trend in brightness. The latter is seen in the \textit{K2} light curves of each of the 8 stars within 5\thinspace arcmin distance from WASP-104. This suggests that the gradual dimming is not astrophysical, and is possibly caused by the imperfect modelling of the background brightness by the Science Operations Center's calibration pipeline.

We used the flattened and normalized light curve for all aspects of the analysis presented in this paper, except for the rotational modulation analysis in Section~5 where we used the fluxed version of the light curve.

\section{REFINEMENT OF SYSTEM PARAMETERS}

We used the Markov chain Monte Carlo (MCMC) procedure presented in \citet{CollierCameron07}, \citet{Pollacco08} and \citet{Anderson15} to obtain the planetary and stellar parameters. With this MCMC procedure we simultaneously analysed the flattened, normalized \textit{K2} transit light curve and the radial velocity (RV) measurements provided by the discovery paper \citep{Smith14}, namely 10 and 11 out-of-transit RV measurements from CORALIE \citep{Queloz00} and SOPHIE \citep{Bouchy09}, respectively. We accounted for stellar limb darkening using a four-parameter law, with coefficients calculated for the \textit{K2} bandpass and tabulated in \citet{Sing10}. We interpolated the limb-darkening coefficients initially using the stellar metallicity ([Fe/H] = $+0.32\pm0.09$) and stellar surface gravity ($\log g = 4.5\pm0.2$) from \citet{Smith14}, and interpolated them at each MCMC step with the latest stellar effective temperature ($T_{\rm eff}$). $T_{\rm eff}$ was used as a free fitting parameter but constrained with a Gaussian prior set at the spectroscopic $T_{\rm eff} = 5450\pm130$\thinspace K from \citet{Smith14}.

After the initial MCMC run, we used the best-fitting transit parameters along with the spectroscopic stellar effective temperature and metallicity given in the discovery paper as inputs to estimate the stellar mass and age with the {\scriptsize{BAGEMASS}} tool \citep{Maxted15}. We used this refined stellar mass estimate in consecutive MCMC runs and noted that the derived system parameters converged already after one such iteration.

Using equation (1) of \citet{Jackson08} and adopting their best-fitting stellar and planetary tidal dissipation parameters of $10^{5.5}$ and $10^{6.5}$, respectively, we estimated the circularization time-scale of WASP-104b as 68\thinspace Myr. Due to this very short circularization time-scale, we imposed a fixed circular orbit in the main MCMC analysis and estimated the eccentricity upper limit in a separate MCMC run where the eccentricity was fitted as a free orbital parameter.

To refine the transit ephemeris, we performed another MCMC run with all the available additional transit photometry from the discovery paper \citep{Smith14}. Beside the \textit{K2} light curve presented in this paper, we included the discovery WASP photometry \citep{Pollacco06}, four transit light curves obtained by TRAPPIST \citep{Jehin11} and two light curves by the Euler Telescope \citep{Lendl12}. Due to the different filters used to obtain the additional light curves, we sourced the four-parameter limb-darkening coefficients for appropriate filters from \citet{Claret00} and \citet{Claret04} and interpolated them in the same way as with the \textit{K2} data. Adding the much-higher photometric precision \textit{K2} light curve to the datasets used by \citet{Smith14} extended the photometric baseline of transits from 4.1 to 8.4 years and reduced the uncertainty of the orbital period by a factor of 19.

We show in Figure~2 the \textit{K2} transit light curve and the best-fitting transit model. The planetary system parameters given in Table~1 agree within 2$\sigma$ and generally have smaller uncertainties than the values presented in \citet{Smith14}.

\begin{figure}
\includegraphics[width=8.5cm]{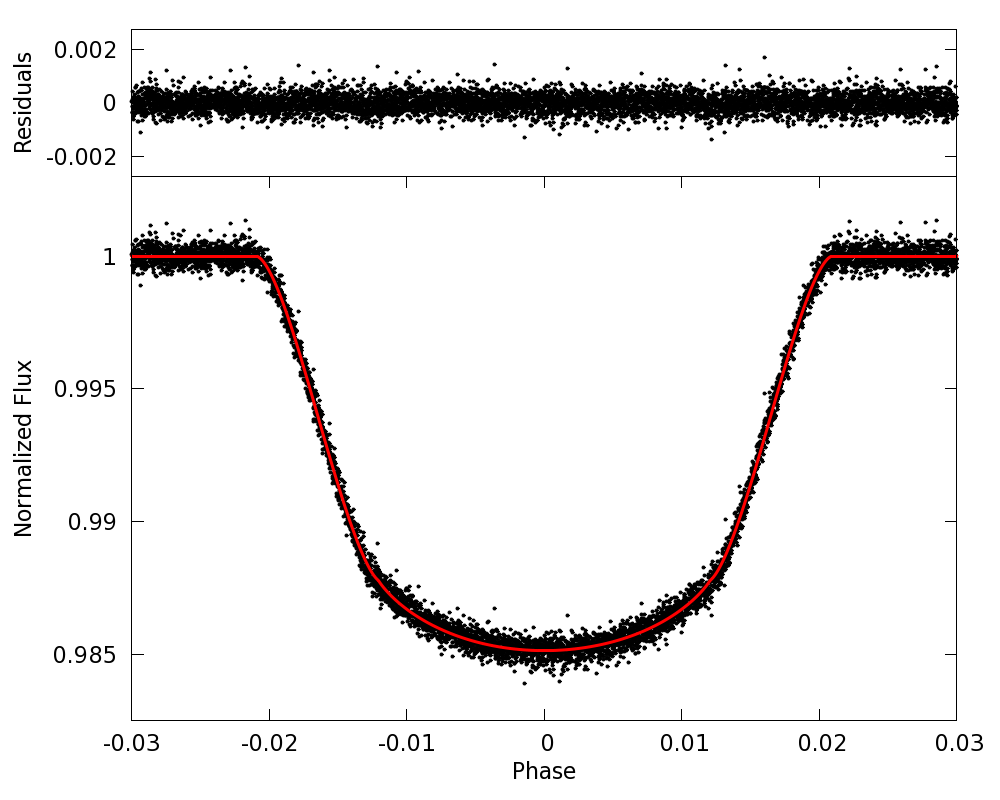}
\caption{Phase-folded \textit{K2} light curve of WASP-104. The red line is the best-fitting MCMC transit model. Shown in the upper panel are the residuals from the transit model.}
\end{figure}

\begin{table}
\centering
\begin{minipage}{8.5cm}
\caption{MCMC system parameters for WASP-104 and WASP-104b}
\begin{tabular}{lccc}
\hline\hline
Parameter&Symbol&Value&Unit\\
\hline
Transit epoch&$t_{\rm 0}$&$2457935.0702321$&BJD$_{\rm TDB}$\\
&&$\pm0.0000086$&\\
Orbital period&$P_{\rm orb}$&1.75540636&days\\
&&$\pm0.00000014$&\\
Area ratio&$(R_{\rm p}/R_{\star})^{2}$&$0.014641\pm0.000020$&...\\
Transit width&$t_{\rm 14}$&$0.072827\pm0.000046$&days\\
Ingress and egress&$t_{\rm 12}$, \textit{t}$_{\rm 34}$&$0.015364\pm0.000079$&days\\
\quad duration&&&\\
Impact parameter&\textit{b}&$0.7278\pm0.0016$&...\\
Orbital inclination&\textit{i}&$83.612\pm0.026$&$^{\circ}$\\
Orbital eccentricity&\textit{e}&0 (adopted)&...\\
&&$<$0.030 at $2\sigma$&\\
Orbital separation&\textit{a}&$0.0286\pm0.00047$&au\\
Stellar mass&$M_{\star}$&$1.011\pm0.050$&$M_\odot$\\
Stellar radius&$R_{\star}$&$0.940\pm0.016$&$R_\odot$\\
Stellar density&$\rho_{\star}$&$1.2178\pm0.0070$&$\rho_\odot$\\
Stellar surface&$\log g_{\star}$&$4.4963\pm0.0074$&cgs\\
\quad gravity&&&\\
Planet mass&$M_{\rm p}$&$1.311\pm0.053$&$M_{\rm Jup}$\\
Planet radius&$R_{\rm p}$&$1.106\pm0.019$&$R_{\rm Jup}$\\
Planet density&$\rho_{\rm p}$&$0.969\pm0.028$&$\rho_{\rm Jup}$\\
Planet surface&$\log g_{\rm p}$&$3.390\pm0.010$&cgs\\
\quad gravity&&&\\
Planet equilibrium&$T_{\rm p}$&$1507\pm39$&K\\
\quad temperature$^{a}$&&&\\
Isochronal age&$\tau_{iso}$&$3.5\pm2.4$&Gyr\\
\quad estimate&&&\\
\textit{K2} limb-darkening&$a_{\rm 1}$, $a_{\rm 2}$&0.693, $-0.426$&...\\
\quad coefficients&$a_{\rm 3}$, $a_{\rm 4}$&0.991, $-0.486$&...\\
\hline
\end{tabular}
\begin{itemize}[leftmargin=0.35cm]
\setlength\itemsep{0cm}
\item[$^{a}$]Planet equilibrium temperature is based on assumptions of zero Bond albedo and complete heat redistribution.
\end{itemize}
\end{minipage}
\end{table}

\section{NO TTV OR TDV}

The time intervals between successive transits and their durations are always the same for an unperturbed planet. However, the transiting planet can exchange energy and angular momentum with a third body. This gravitational interaction causes short-term oscillations of semi-major axes and eccentricities, which may result in measurable TTVs (e.g. \citealt{Holman10}) or TDVs (e.g. \citealt{Nesvorny13}). Largest variation amplitudes are expected for planets near low-order resonance orbits with perturbing objects \citep{Lithwick12}. The detection of such variations allows the determination of orbital periods and masses of additional objects in planetary systems \citep{Holman05}.

We measured the TTVs and TDVs of WASP-104b by modelling each transit in the short-cadence \textit{K2} light curve individually and subtracting the measured individual transit timings and durations from the best-fitting ephemeris given in Table~1. The MCMC procedure of transit modelling was similar to that in Section~3, except that we fitted only the transit timings and durations while keeping other observables fixed at their best-fitting values from Table~1. Under an assumption of white noise distribution around zero, we calculated the $\chi^2$ values as
\begin{equation}
\chi^2=\sum_{i=1}^{45} \frac{(O_i-\langle O\rangle)^2}{(\Delta O_i)^2}\,,
\end{equation}
where $i$ is the transit number, $O_i$ is observed TTV or TDV of the $i^{\rm th}$ transit, $\langle O\rangle$ is the mean of the observed TTVs or TDVs and $\Delta O_i$ is the TTV or TDV uncertainty of the $i^{\rm th}$ transit. We obtain $\chi_{\rm TTV}^2 = 57.4$ and $\chi_{\rm TDV}^2 = 30.3$, for 44 degrees of freedom. Thus, we do not detect any statistically significant periodic signals in either TTVs nor TDVs. We place the semi-amplitude upper limits at 20\thinspace s and 47\thinspace s for TTVs and TDVs, respectively, for periods shorter than 80 days. The upper limits were determined as three times the weighted standard deviations. As an illustration, by using the equations from \citet{Lithwick12}, the obtained TTV upper limit implies the absence of any non-transiting planets within 10\% of the 2:1 resonance circular orbits and masses above 23\thinspace $M_{\rm Earth}$.

\section{ROTATIONAL MODULATION}

Starspots can induce brightness modulations as they are coming and going from the field of view while the star rotates. The periodicity of rotational modulation is therefore indicative of the stellar rotational period (e.g. \citealt{McQuillan13}).

A brightness modulation with a time-scale of tens of days is visible even in Figure~1. The modulation can be seen much more clearly in Figure~3 where we show the light curve at a larger scale with transits and a linear dropping brightness trend removed, and binned by a factor of 50 to reduce white noise. The modulation is not correlated with the position of the target on the detector and is not present in the \textit{K2} light curves of other nearby stars, which indicates the modulation to be of astrophysical origin. The most likely cause of the observed modulation is the presence of starspots on the surface of the rotating host star.

\begin{figure}
\includegraphics[width=8.5cm]{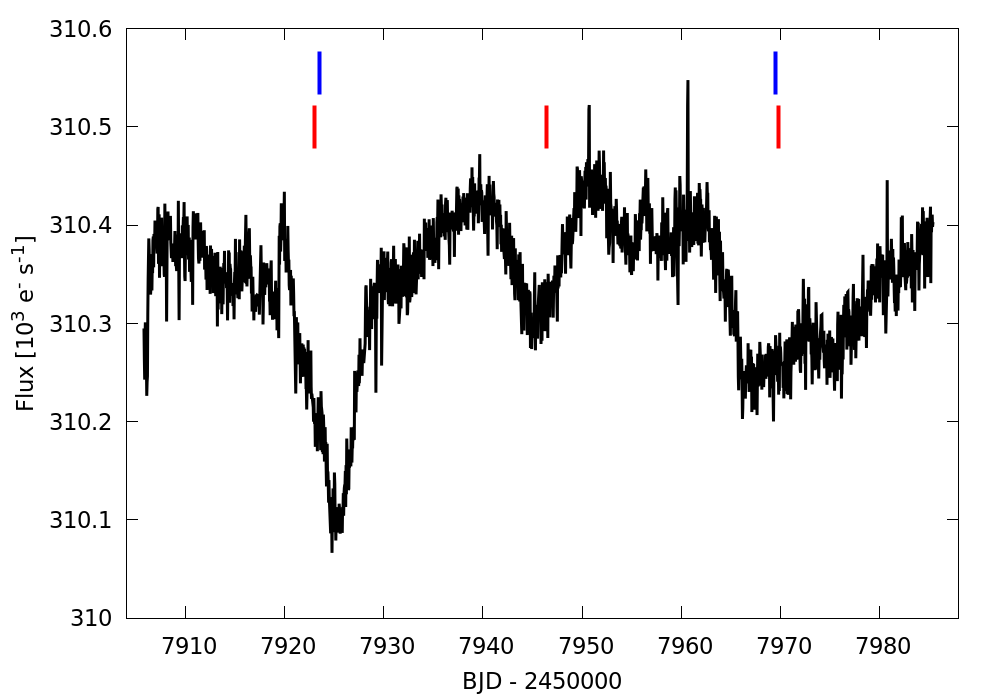}
\caption{Binned \textit{K2} light curve with transits and a linear dropping brightness trend removed. The rotational modulation is clearly visible, with ticks indicating the potential minima of either the 23- (red) or the 46-day (blue) periodicity.}
\end{figure}

To determine the period of the modulation, we calculated the Lomb--Scargle periodogram (see Figure~4) of the light curve shown in Figure~3. The two highest peaks are at $46^{+15}_{-7}$ and $23.4\pm2.7$ days. The rotational period is likely to be one of these, though the \textit{K2} data do not cover enough cycles to be sure. The periodicity near 46 days would correspond to a rotational period longer than half of the 80-day baseline and is therefore not reliable. If the true period were 23 days then peaks at 46 and 10.5 days would be aliases at twice and half of the rotational period.

\begin{figure}
\includegraphics[width=8.5cm]{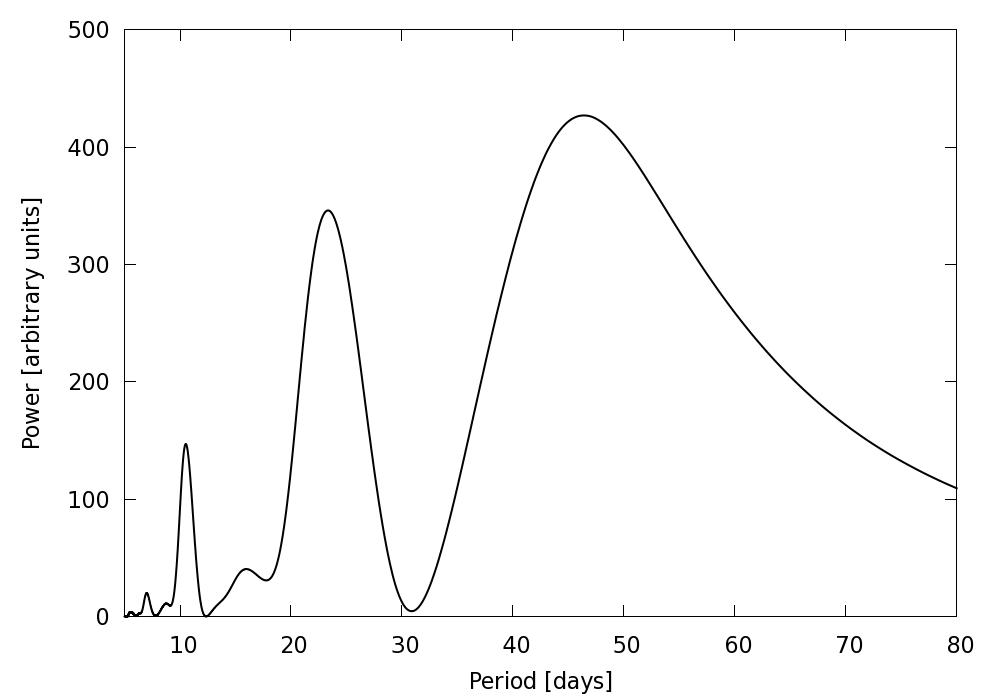}
\caption{Lomb--Scargle periodogram showing a probable rotational period of $23.4\pm2.7$ or $46^{+15}_{-7}$ days.}
\end{figure}

The non-detection of such rotational modulation by \citet{Smith14} is unsurprising, given that their semi-amplitude upper-limit was 4\thinspace mmag. Their detection threshold was an order of magnitude above our detection from the \textit{K2} data which reveal a modulation with a semi-amplitude of about 400\thinspace ppm.

\citet{Smith14} provided a stellar projected rotational velocity of $0.4\pm0.7$\thinspace km\thinspace s$^{-1}$, which for the stellar radius given in Table~1 yields a 1-$\sigma$ lower limit of rotational period of 43 days assuming that the star rotates edge on. This slow stellar rotation is compatible with the potential 46-day rotational period. Alternatively, the 23-day period would be in agreement with the projected rotational period only if the inclination of the stellar rotational axis were smaller than $\sim$33$^{\circ}$.

\section{NO STARSPOT OCCULTATIONS}

Starspot occultations are the in-transit brightening events that occur whenever a starspot is occulted by a transiting planet \citep{Silva03}. The same starspot may be occulted repeatedly in several transits (e.g. \citealt{Tregloan13}), or different starspots may be occulted at similar preferential transit phases (e.g. \citealt{Sanchis11}). Starspot occultation events may be used for an independent and precise measurement of the stellar rotational period \citep{Silva08} and a measurement of the misalignment angle between stellar rotational and planet's orbital axis \citep{Nutzman11}.

\citet{Mocnik16a} have shown that detecting starspot occultation events in the \textit{K2} datasets is possible despite the reduced pointing stability of the spacecraft.

We subtracted the best-fitting transit model from the short-cadence \textit{K2} light curve of WASP-104 and searched by eye for any in-transit starspot occultation events. As we found no occultations, we set an occultation amplitude upper limit to 840\thinspace ppm, equal to twice the highest in-transit standard deviation.

The presence of a rotational modulation (see Section~5) suggests that starspots should be present. However, the amplitude of the modulation is much lower than in other systems that show starspot occultation events. For example, Qatar-2 has a rotational modulation with an amplitude of about 2\% \citep{Mocnik17b}, which is 25 times higher than that in WASP-104. If the starspot occultations were also 25 times smaller than those seen in Qatar-2 then they would not be observable.

However, if the rotational modulation in WASP-104 were caused by a single spot, that was completely occulted in transit, then we might expect a starspot occultation with an amplitude comparable to the amplitude of the rotational modulation (800\thinspace ppm), which would be marginally detectable. On the other hand, the likelihood of the transit chord passing over a single spot is low.

If the planet's orbit is aligned, then the transit chord could be at a different latitude than stellar active regions. Indeed, this would be expected given that the impact parameter of the planet is large at 0.73, and knowing that most sunspots occur within $40^{\circ}$ of the solar equator \citep{Mandal17}. If, instead, the planet's orbit is misaligned, then over 45 consecutive transits, the transit chord samples many more latitudes and so the likelihood that it crosses a spot is much higher.

Thus, without knowing the alignment of the orbit, we cannot draw firm conclusions from the absence of detectable starspot occultations. It may therefore be worth obtaining Rossiter--McLaughlin observations of this system to measure the alignment.

\section{PHASE-CURVE MODULATION}

For the phase-curve analysis we used the \textit{K2} light curve which was flattened and normalized as described in Section~2. Flattening was needed to remove any low-frequency brightness variability such as the rotational modulation (see Section~5). As a test to ensure that the flattening procedure did not affect also the phase-curve modulation, we injected a suite of phase-curve signals prior to flattening and successfully recovered them after the flattening.

Figure~5 shows the final phase curve with a binning of 50 bins. The binning factor was chosen by trial and error to find a good compromise between lowering the white noise and retaining the phase-resolution. As can be seen even by eye in Figure~5, the phase curve exhibits a signal resembling an ellipsoidal modulation. We ran an MCMC procedure to model the phase curve with three phase-curve modulation components \citep{Mazeh10}:

\begin{align}
&F_{\rm ell} = -A_{\rm ell}\cos\left(\frac{2\pi}{P_{\rm orb}/2}t\right)\ ,\\
&F_{\rm Dop} = A_{\rm Dop}\sin\left(\frac{2\pi}{P_{\rm orb}}t\right)\ ,\\
&F_{\rm ref} = -A_{\rm ref}\cos\left(\frac{2\pi}{P_{\rm orb}}t\right) + F_{\rm sec}\ ,
\end{align}
where $F_{\rm ell}$, $F_{\rm Dop}$, and $F_{\rm ref}$ are ellipsoidal, Doppler beaming, and reflection modulation components in the normalized phase curve, respectively. $A_{\rm ell}$, $A_{\rm Dop}$, and $A_{\rm ref}$ are the corresponding semi-amplitudes, $P_{\rm orb}$ is the orbital period, and $t$ is time from mid-transit. $F_{\rm sec}$ is a simplified secondary eclipse signal, whose depth equals to the reflection amplitude and the duration is the same as for the transit given in Table~1. Because the thermal emission is expected to be small compared to the reflected light in the \textit{K2} optical bandpass, in this paper we refer to the combined signal from the planetary reflection and its thermal emission simply as reflection modulation.

\begin{figure}
\includegraphics[width=8.5cm]{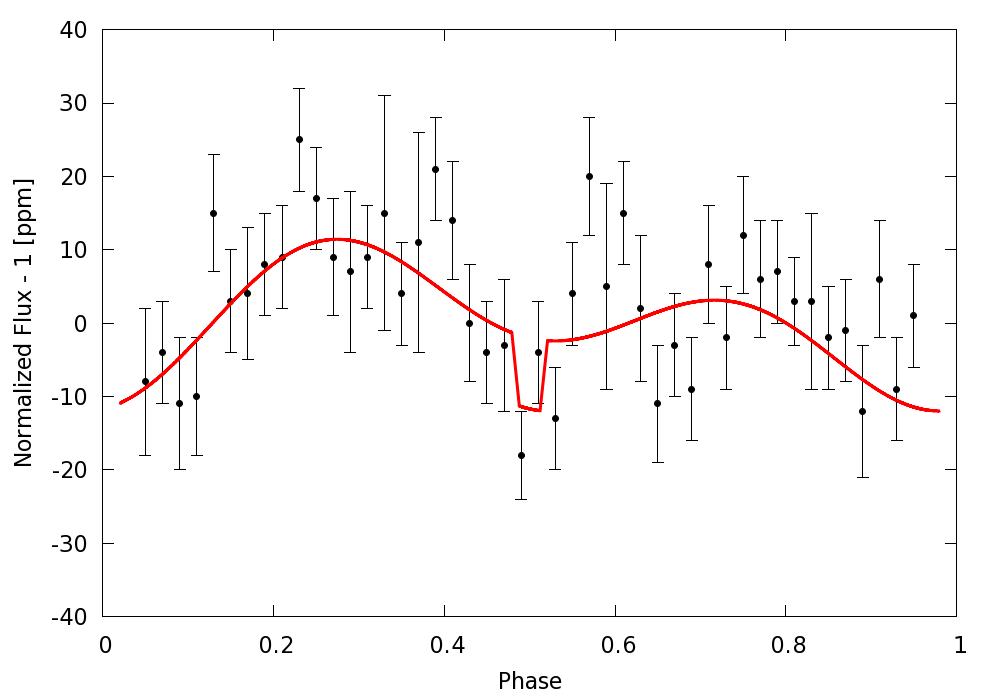}
\caption{Binned phase curve of WASP-104. The best-fitting MCMC phase-curve model is shown with a red line. The ellipsoidal modulation component is detected with high significance, whereas the detections of Doppler and reflection components are tentative.}
\end{figure}

The best-fitting MCMC phase-curve model is shown with a red line in Figure~5 and corresponds to the ellipsoidal, Doppler and reflection signals with semi-amplitudes of $6.9\pm2.2$, $4.2\pm1.9$, and $4.8\pm2.1$\thinspace ppm, respectively. Modelling only the ellipsoidal modulation yields a change in the Bayesian information criterion (BIC) of 13.6, while adding first Doppler and then also the reflection modulations further increases the $\Delta {\rm BIC}$ to 16.9 and 18.8, respectively. This implies that the detection of ellipsoidal modulation is strong, while the Doppler and reflection modulation detections are tentative.

As a test, we repeated the analysis using only the first and then only the second half of the light curve, and found that the modelled phase curve modulations are in agreement with the results obtained from analysing the full light curve. We also performed the same analysis on the much-noisier publicly available long-cadence light curve reduced with the K2SFF procedure \citep{Vanderburg14} and somewhat noisier short-cadence light curve detrended with K2SC \citep{Aigrain16}, optimized for short-cadence data \citep{Mocnik17a}. In both cases we detect ellipsoidal modulation in agreement with our detection, however, we cannot confirm the Doppler and reflection modulations with these two light curves.

Using the system parameters from Table~1 and equations (7--10) of \citet{Mazeh10}, we estimate the theoretical semi-amplitudes of ellipsoidal, Doppler, and reflection modulations to be 5.7, 2.7, and 
$330A_{\rm g}$\thinspace ppm, respectively, where $A_{\rm g}$ is the geometric albedo of the planet. The semi-amplitudes of our ellipsoidal and Doppler detections agree with the theoretically predicted values. For the same to be true also for reflection, the $A_{\rm g}$ would have to be of the order of one percent. Because the detection of the reflection modulation is tentative, we provide here only the 2-$\sigma$ upper limit for the visual geometric albedo of 0.03.

At such small reflectivity the thermal emission may contribute significantly to the detected combined phase-curve signal. Thermal emission is strongest for planets with large day-night temperature contrasts, resulting in an emission phase-curve component resembling reflection. Weakest thermal emission is produced by planets with low day-night temperature contrasts, with a flat emission phase-curve signal \citep{Heng13}. By using the refined system parameters, we calculated that the planet's thermal emission in the \textit{K2} bandpass contributes a minimum of 1\thinspace ppm deeper occultation depth and no sinusoidal reflection-like phase-curve signal in the case of complete heat redistribution in the planet's atmosphere. In the case of no heat redistribution and tidal locking, the emission phase-curve signal would superimpose with reflection by adding 3.3\thinspace ppm to its semi-amplitude. Therefore, the true geometric albedo is likely to be significantly lower than the upper limit given above. To break the degeneracy between reflection and emission, we would require a set of phase-curve observations in another wavelength region, preferably in the infrared where the planetary emission component is much stronger.

\section{NO ADDITIONAL TRANSITING PLANETS}

\citet{Kovacs02} introduced a box-fitting least squares (BLS) algorithm to detect periodic transit-like signals in photometric datasets. We searched for any additional transiting planets with the BLS algorithm in our flattened and normalized \textit{K2} light curve, from which we removed the data points within 0.025 phase from transit mid-points of WASP-104b. This was done by using the online BLS periodogram service provided by the NASA Exoplanet Archive\footnote{\mbox{https://exoplanetarchive.ipac.caltech.edu/cgi-bin/Pgram/}nph-pgram}. We then converted the obtained BLS signal residuals into estimated transit depths of potential transiting planets \citep{Kovacs02}:

\begin{equation}
\delta = \frac{SR}{\sqrt{r(1-r)}}\ ,
\end{equation}
where $\delta$ is the transit depth, $SR$ is BLS signal residual, and $r$ is the relative time spent in transit, which we approximated with the transit phase-width a potential planet would have at a particular orbital period.

We found no statistically significant periodogram peaks in the period region 0.5--30 days, and set a transit depth upper limit to 110\thinspace ppm, which equals the highest peak in the residual transit-depth periodogram.

\section{CONCLUSIONS}

WASP-104 was observed by the \textit{K2} in the short-cadence mode during the observing Campaign 14. By analysing these data we refined the system parameters and searched for TTVs, TDVs, rotational modulations, starspot occultations, phase-curve modulations, and additional transiting planets.

We detect the rotational modulation with a probable rotational period of 23 or 46 days. Despite the apparent presence of starspots, we did not detect any starspot occultation events, possibly due to the large impact parameter of the transiting planet or because the occultations did not exceed our detection threshold.

WASP-104 is, to the best of our knowledge, only the third transiting planetary system with detected phase-curve modulation from the \textit{K2} mission (after Qatar-2 \citep{Mocnik17b,Dai17} and K2-141 \citep{Malavolta18}). We unequivocally detect ellipsoidal modulation with a semi-amplitude of 7\thinspace ppm, in agreement with the theoretically expected value. We also tentatively detect Doppler beaming and reflectional modulations. The latter yields a conservative upper limit for the planet's visual geometric albedo of 0.03, lower than the reflectance of charcoal \citep{Ascough10}. The very low albedo rules out any highly reflective clouds in the WASP-104b's atmosphere.

TrES-2b is one of very few hot Jupiters at least as dark as WASP-104b. \citet{Kipping11b} have measured its visual geometric albedo to be $0.025\pm0.007$ if the detected reflectional modulation in the \textit{Kepler} data was caused entirely by reflection, and even lower than 1\% after taking into account their thermal emission model. Another example is HAT-P-7b, with a visual geometric albedo $\lesssim$0.03, based on the detection of the secondary eclipse in the \textit{Kepler} light curve \citep{Morris13}.

In general, hot Jupiters exhibit a large range of visual geometric albedos (e.g. \citealt{Sheets17}), depending on their temperature which controls the cloud properties \citep{Sudarsky00}. Typical visual geometric albedos of hot Jupiters are of the order of 0.1 \citep{Schwartz15} and are statistically lower than for hot super-Earths \citep{Demory14} and Neptunes \citep{Sheets17}. According to the atmospheric models, the lower albedos may be attributed to the presence of alkali metals as well as TiO and VO in hot-Jupiter atmospheres, which causes significant absorption in the visual wavelengths \citep{Demory11}.

\acknowledgements{We gratefully acknowledge the financial support from the Science and Technology Facilities Council, under grants ST/J001384/1, ST/M001040/1 and ST/M50354X/1. This paper includes data collected by the \textit{K2} mission. Funding for the \textit{K2} mission is provided by the NASA Science Mission directorate. This work made use of PyKE \citep{Still12}, a software package for the reduction and analysis of \textit{Kepler} data. This open source software project is developed and distributed by the NASA Kepler Guest Observer Office. This research has made use of the NASA Exoplanet Archive, which is operated by the California Institute of Technology, under contract with the National Aeronautics and Space Administration under the Exoplanet Exploration Program.}

\bibliographystyle{apj}
\bibliography{bibliography}

\end{document}